 \documentclass[final,1p,times]{elsarticle}

\usepackage{amssymb}

\usepackage{amsmath,amssymb,amsfonts}
\usepackage{algorithmic}
\usepackage{graphicx}
\usepackage{subcaption}
\usepackage{textcomp}
\usepackage{multicol}
\usepackage[section]{placeins}
 \usepackage{float}
\usepackage{stfloats}

\usepackage{lipsum}
\makeatletter
\def\ps@pprintTitle{%
 \let\@oddhead\@empty
 \let\@evenhead\@empty
 \def\@oddfoot{}%
 \let\@evenfoot\@oddfoot}
\makeatother

\journal{Nuclear Physics B}

\begin{document}

\begin{frontmatter}


\title{An Artificial Neural Network for Gait Analysis to Estimate Blood Alcohol Content Level}

 \author{Pedram Gharani}
 \ead{peg25@pitt.edu}
 \address{School of Computing and Information, University of Pittsburgh}
\author{Brian Suffoletto, M.D.}
\ead{suffbp@upmc.edu}
\address{Department of Emergency Medicine, School of Medicine, University of Pittsburgh}

\author{Tammy Chung, Ph.D.}
\ead{chungta@upmc.edu}
\address{Department of Psychiatry, School of Medicine, University of Pittsburgh}

\author{Hassan Karimi, Ph.D.}
\ead{hkarimi@pitt.edu}
 \address{School of Computing and Information, University of Pittsburgh}
\begin{abstract}
Impairments in gait occur after alcohol consumption, and, if detected in real-time, could guide the delivery of "just-in-time" injury prevention interventions. We aimed to identify the salient features of gait that could be used for estimating blood alcohol content (BAC) level in a typical drinking environment. We recruited 10 young adults with a history of heavy drinking to test our research app. During four consecutive Fridays and Saturdays, every hour from 8pm to 12am, they were prompted to use the app to report alcohol consumption and complete a 5-step straight-line walking task, during which 3-axis acceleration and angular velocity data was sampled at a frequency of 100 Hz. BAC for each subject was calculated. From sensor signals, 24 features were calculated using a sliding window technique, including energy, mean, and standard deviation. Using an artificial neural network (ANN), we performed regression analysis to define a model determining association between gait features and BACs. 70\% of data was used as a training dataset, and the results were tested and validated using the rest of samples. We evaluated different training algorithms for the neural network and the result showed that a Bayesian regularization neural network (BRNN) was the most efficient and accurate. Analyses support the use of the tandem gait task paired with our approach to reliably estimate BAC based on gait features. Results from this work could be useful in designing effective prevention interventions to reduce risky behaviors during periods of alcohol consumption.
\end{abstract}

\begin{keyword}


Neural Network, Bayesian regularization neural network (BRNN), Blood Alcohol Content (BAC), Feature extraction
\end{keyword}
\end{frontmatter}



\section{Introduction}
Acute alcohol intoxication is associated with numerous health risks. For example, impaired driving due to alcohol was implicated in 31\% of the 33,000 deaths from motor vehicle accidents in the US in (NHTSA, 2015 \citep{NCSAnalysis2015}). These consequences largely stem from alcohol’s impairing effects on psychomotor performance \citep{christoforou2013reaction}. Compounding this risk are impaired decision-making \citep{steele1990alcohol} and lack of awareness of the degree of alcohol-related impairments during drinking episodes \citep{morris2014perceived}.  Strategies to measure alcohol-related psychomotor impairments and provide real-time feedback to individuals could deter involvement in activities that require psychomotor function (i.e. driving), thus reducing likelihood of injury \citep{shults2001reviews}.  

One measure of psychomotor performance that is sensitive to alcohol is walking (gait). Gait requires coordination of multiple sensory and motor systems. Research in a controlled laboratory setting has shown that alcohol affects both postural stability \citep{nieschalk1999effects} and gait \citep{jansen1985gait}. Although law enforcement professionals have used subjective performance on a heel-to-toe tandem gait task as a field sobriety test for years, there is no current process to objectively measure aspects of gait during drinking occasions.  The rapid growth of smartphone ownership \citep{anderson2015pew} suggests that these devices could be useful to objectively measure gait impairment during drinking episodes. Although a couple of small field studies have used smartphone accelerometers and gyroscopes to detect gait abnormalities during alcohol consumption \citep{Arnold2015}, \citep{Aiello2016}, none has determined the association of gait features with blood alcohol content (BAC) levels.

The purpose of this work was to identify the features of movement patterns (gait) that can be measured through a smartphone’s 3-axis accelerometer and gyroscope that could be used for estimating the BAC level in a typical drinking environment. To accomplish this aim, we designed an iPhone app (DrinkTRAC) to collect smartphone sensor-based data on gait (3-axis accelerometer, gyroscope, magnetometer) and ecological momentary assessment (EMA) measures of self-reported number of drinks consumed each hour, from 8pm to 12am, during weekend evenings (Fridays and Saturdays). We did not collect detailed data on drinking in the hours prior to 7pm start (1 hour prior to 8pm), which could have under-estimated BAC in some cases. However, based on prior research in a similar cohort of young adults \citep{bae2017mobile}, we found that less than 15\% of drinking occasions start before 7pm. We enrolled 10 young adults with a history of heavy drinking in a repeated-measures study to provide smartphone sensor and self-report data over a period of four consecutive weeks. We used a Bayesian regularized neural network (BRNN) to perform regression analysis to estimate BAC. Results from this work could be useful in designing effective prevention interventions to reduce risky behaviors during periods of alcohol intoxication. 
In this paper, we propose a novel approach for analyzing the movement patterns of people and develop a supervised learning model to associate their gait anomalies with BAC levels. The gait features are captured using smartphone-based inertial measurements, using accelerometer, gyroscope, and magnetometer sensors. By exploring the aspects of gait and extracting their salient features, we input them to a supervised machine learning technique. Thus, the primary goal of this work is to develop and implement a gait analyzing system, using the inertial sensors of smartphones inside a user's pocket, while simultaneously capturing their movement pattern for the detection of alcohol-induced changes in gait patterns. The results of this work provide us with an in-depth understanding of the spatiotemporal properties of human gait that are affected by alcohol. In summary, the contributions of the paper are:

\begin{itemize}
\item Exploring and identifying gait properties and extracting some features from gait signals measured by smartphone sensors that could estimate BAC values in a typical drinking environment.
\item Comparing different machine-learning techniques to predict BAC values.
\item Demonstrating the feasibility of smartphone sensors measurements in estimating BAC value.
\end{itemize}

The rest of the paper is structured as follows: In Section 2, we explain methods to measure alcohol consumption and the ways in which BAC levels are computed. In Section 3, we describe our method for collecting data through the smartphone-based application. In Section 4, we explain sensor data processing for movement pattern analysis and extracting gait features. We also discuss some details of our network architecture and training procedures. In Section 5, we present the results comparing different training techniques for the artificial neural network and the process of finding the most efficient technique. Furthermore, we evaluate two other regression algorithms. In Section 6, we discuss some related work. Finally, in Section 7, we provide concluding remarks and future directions.

\section{Background: Methods to Measure Alcohol Consumption}
The ability to accurately measure BAC in the real world is vital for understanding the relationship between alcohol consumption patterns and the impairments of normal functioning that occur (such as those related to gait). BACs vary as a function of gender, total amount of alcohol consumed, type of alcohol, time spent drinking, food consumption, body weight and individual differences in absorption and metabolism rates \citep{winek1984blood}. Available methods to measure BAC outside of healthcare facilities include self-reporting and non-invasive monitoring methods i.e., breathalyzer or transdermal alcohol monitors \citep{greenfield2014biomonitoring}. Due to limitations in feasibility of measuring in-vivo BACs using monitors, the majority of scientific literature on alcohol consumption has been based on retrospective self-reports \citep{wechsler2002trends}. In 1932, the first equation to estimate blood-alcohol content from self-reported alcohol consumption was published \citep{widmark1932theoretischen}. Since then, researchers have identified ways to improve the accuracy of that formula, such as modifications primarily in how they adjust body weight to account for gender differences in water content of the body and secondarily, in how the overall dose of alcohol is calculated. One of the most accurate formulas was created by Matthews \& Miller in 1979:

\begin{equation}
eBAC=(\frac{c}{2} \times \frac{GC}{weight})-(\beta 60 \times t)
\end{equation}

Here, BAC is blood alcohol concentration expressed in g/dl; c is the number of standard drinks reported; GC is a gender constant (9.0 for women and 7.5 for men); β60 is the metabolism rate of alcohol per hour (.017 g/dl); and t is the number of hours spent drinking. This formula was found to have a significantly stronger intraclass correlation with breath alcohol concentrations (criterion standard) than did the other equations when measured after an uncontrolled episode of drinking \citep{hustad2005using}. Still, the accuracy of self-reported eBAC values are dependent on a respondents' ability to recall the number of drinks they consumed, knowledge of standardized drink sizes, and the absence of reporting biases due to minimizing sensitive information \citep{babor2000talk}.

“In this study, we chose not to use transdermal alcohol monitors or breathalyzers for several reasons. First, transdermal alcohol monitors (e.g., WrisTAS, SCRAM) are relatively costly to acquire and maintain, which can limit their wide use. Second, transdermal alcohol monitors and breathalyzers involve some burden for participants (e.g., possible minor skin irritation from SCRAM, need to carry breathalyzer) \citep{alessi2017experiences},\citep{simons2015quantifying}, \cite{alessi2013randomized}, \citep{suffoletto2017using}. By comparison, smartphone sensor data, can be collected with relatively low burden, unobtrusively on an individual’s personal phone \citep{suffoletto2017using}.  Third, transdermal alcohol monitors have been found to be less useful in detecting lower drinking quantities, as compared to self-reports, and content readings tend to lag behind consumption by up to several hours \citep{leffingwell2013continuous},\citep{karns2016time}. 

We used retrospective self-reports and the Matthews \& Miller formula to estimate eBAC. To assist with recall, we asked participants to report their number of drinks per hour. The use of experience sampling methods to collect self-reports of alcohol use that is more proximal to drinking occasions can minimize any biases associated with retrospective reporting \citep{muraven2005daily}. To ensure the standardization of drink amounts, the DrinkTRAC app presented participants with a color picture of "standard drink" sizes (based on National Institute on Alcohol Abuse and Alcoholism guidelines \citep{us2005helping}: 12 oz of beer, 5 oz of wine, 1.5 oz of liquor) and asked: "How many standard drinks did you have in the past hour?" with a drop-down menu ranging from 0 to 30. In this way, participants reported alcohol consumption in standard drink units, in order to minimize error in self-report of alcohol consumption. Self-report of alcohol use using EMA has shown validity \citep{shiffman2009ecological}, \citep{Wray2014}. To reduce reporting biases, we used a technology platform to collect sensitive data, which has been shown to be more accurate than in-person reporting \citep{lucas2014s}.

\section{Data Collection}
\subsection{Smartphone application ("DrinkTRAC") for data collection}

This prospective study recruited a convenience sample of young adults who were identified in the emergency department (ED) as reporting past hazardous drinking between February 19 and May 9, 2016. All participants completed informed consent protocols prior to study procedures and were provided with resources for alcohol treatment. 

\subsection{Participants}

Participants were young adults (aged 21–-26 years) who presented to an urban ED. A total of 28 medically stable ED patients who were not seeking treatment for substance use, not intoxicated, and who were going to be discharged to home, were approached by research staff. Among those eligible to be approached, 23 patients provided consent to complete an alcohol use severity screen. Those who reported recent hazardous alcohol consumption based on an Alcohol Use Disorder Identification Test for Consumption (AUDIT-C) score of $\geq$3 for women or $\geq$4 for men \citep{babor1992guidelines} and who drank primarily on weekends were eligible for participation. We excluded those who reported any medical condition that resulted in impaired thinking or memory or gait, those who reported past treatment for alcohol use disorder, and those without an iOS phone. A total of 10 participants met the study enrollment criteria and uploaded the DrinkTRAC app to their phone. We instructed participants to refrain from any non-drinking substance use (excluding cigarette use) during the sampling days. We also informed participants that they would receive \$10 for completing the baseline survey and app-based tasks in the ED, \$10 for completing the exit survey at four weeks, and \$1 per completed EMA (up to an additional \$40). Table \ref{table1} shows the results for sample descriptive statistics.

\begin{table}[]
\centering
\caption{Sample descriptive statistics}
\label{table1}
\begin{tabular}{lllc}
\hline
\multicolumn{3}{l}{Characteristics} & N=10 \\ \hline
\multicolumn{3}{l}{Age in years, mean (SD)} & 23.1 (2.6) \\
\multicolumn{3}{l}{Female, n (\%)} & 7 (70\%) \\
\multicolumn{3}{l}{Race, n (\%)} &  \\
 & \multicolumn{2}{l}{African American} & 2 (20\%) \\
 & \multicolumn{2}{l}{White} & 6 (60\%) \\
 & \multicolumn{2}{l}{Other} & 2 (20\%) \\
\multicolumn{3}{l}{Hispanic Ethnicity, n (\%)} & 1 (10\%) \\
\multicolumn{3}{l}{Education, n (\%)} &  \\
 & \multicolumn{2}{l}{Some college} & 5 (50\%) \\
 & \multicolumn{2}{l}{College graduate or,post-graduate} & 5 (50\%) \\
\multicolumn{3}{l}{Employment, n (\%)} &  \\
 & \multicolumn{2}{l}{For wages} & 7 (70\%) \\
 & \multicolumn{2}{l}{Student} & 3 (30\%) \\
\multicolumn{3}{l}{Married, n (\%)} & 1 (10\%) \\
\multicolumn{3}{l}{Alcohol Use Severity (AUDIT-C score), mean (SD)} & 5 (1.3) \\
\multicolumn{3}{l}{Weight in pounds, mean (SD)} & 179 (35) \\ \hline
\end{tabular}
\end{table}

\subsection{Smartphone Application Design}
The DrinkTRAC app was developed using Apple's ResearchKit platform, as it allowed for convenient and professional-appearing modular builds that incorporated timed psychomotor tasks. Baseline survey questions included socio-demographic measures and severity of alcohol use. The app then presented participants with EMA, including two questions (cumulative number of drinks consumed and perceived intoxication) followed by psychomotor tasks, including a 5--step tandem gait task. The research associate was present to ensure understanding and to observe compliance with instructions on the initial trial of the app's tasks, which were conducted in the ED.

Over four consecutive Fridays and Saturdays, every hour from 8pm to 12am, participants were sent an electronic notification to log in to the DrinkTRAC app and complete the EMA. We chose to sample data on weekend evenings, given that this is a time when young adults typically drink alcohol \citep{del2004up}. We collected EMAs hourly from 8pm to 12am on those nights, with an intention of capturing both the ascending and descending limbs of alcohol intoxication. We used fixed hourly assessment times, given that they would provide a predictable framework for participants and would allow us to more easily calculate eBAC changes over the course of the evening. Given the deleterious effect of alcohol on memory, we chose to collect drinks consumed since last report as opposed to cumulative drinks over an entire drinking occasion We designed the tandem gait task to take less than 45 seconds to optimize completion and reduce potential for disruptions that could interfere with task performance. Basic text instructions were given prior to the tandem gait task, and when the task was completed, participants were presented with a figure of their completion rates for the day.

Estimated Blood Alcohol Concentration. We calculated eBAC during each hour when data was available using the aforementioned formula, created by Matthews and Miller \citep{matthews1979estimating}. Estimates produced by this formula correlate with breath alcohol concentration and were found to perform best, relative to estimates from other commonly used eBAC formulas \citep{hustad2005using}. When drinks consumed in any prior hours were missing, we assumed no drinks were consumed during that period. When drinks had been consumed in prior hours, we incorporated those drinks into BAC calculations (with alcohol clearance taken into account). As shown in related research using the same data set \citep{suffoletto2017can}, participants completed 32\% of EMA. Higher rates of missing EMA data occurred later in the evening and over time in the study. Within the 128 completed EMA, we captured 38 unique drinking episodes, with each participant reporting at least 3 drinking episodes. Almost half of the EMA (n=60, 46.9\%) were completed either prior to drinking or on non-drinking evenings, 55 EMA (43.0\%) were completed on the ascending eBAC limb, and 13 (10.1\%) were completed on the descending eBAC limb. However, there were a number of occasions with missing data on alcohol consumption in hours prior to a given hour were missing and thus assumed to be zero. This may have resulted in under-estimations of BAC. On a drinking day, participants reported consuming a mean of 3.6 (SD=2.2; range: 1-10 drinks). The mean eBAC was 0.04 (SD=0.05), with a peak of 0.23.

\subsection{Inertial data acquisition during tandem gait task}
\begin{figure}[H]
\begin{center}
\includegraphics[scale=.5]{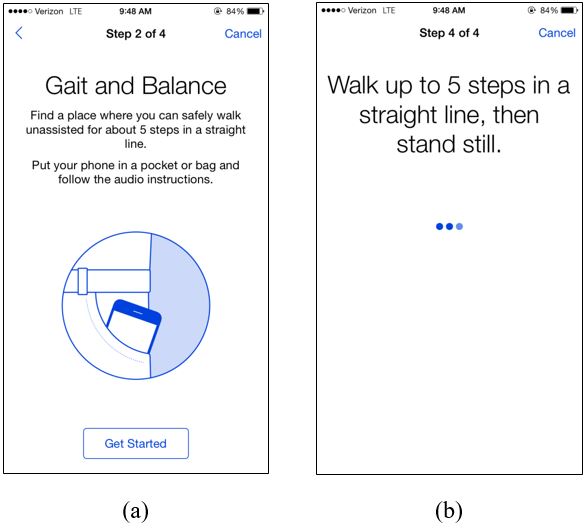}
\caption{DrinkTRAC app screenshots of the tandem gait task}
\label{fig1}
\end{center}
\end{figure}

In the tandem gait task, participants were instructed to walk in a straight line for 5 steps. We advised participants not to continue if they felt that they could not safely walk 5 steps in a straight line unassisted. If participants clicked "next", they were shown a picture of a phone in a front pocket and told: "Find a place where you can safely walk unassisted for about 5 steps in a straight line", followed by the text: "Put the phone in a pocket or bag and follow the audio instructions. If you do not have somewhere to put the phone, keep it in your hand". When the participant clicked "Get Started", the app displayed a timer and played an audio recording of a voice counting down from 5 to 1. If the audio option was turned on, participants heard "Walk up to 5 steps in a straight line, then stand still." We recorded the acceleration with gyroscope sensors embedded in the phone to collect 3-axis acceleration and angular velocity at a sampling frequency of 100 Hz for 30 seconds Figure \ref{fig1}(a) and (b) are DrinkTRAC app screen shots of the tandem gait task

\begin{figure}[H]
\begin{center}
\includegraphics[scale=.295]{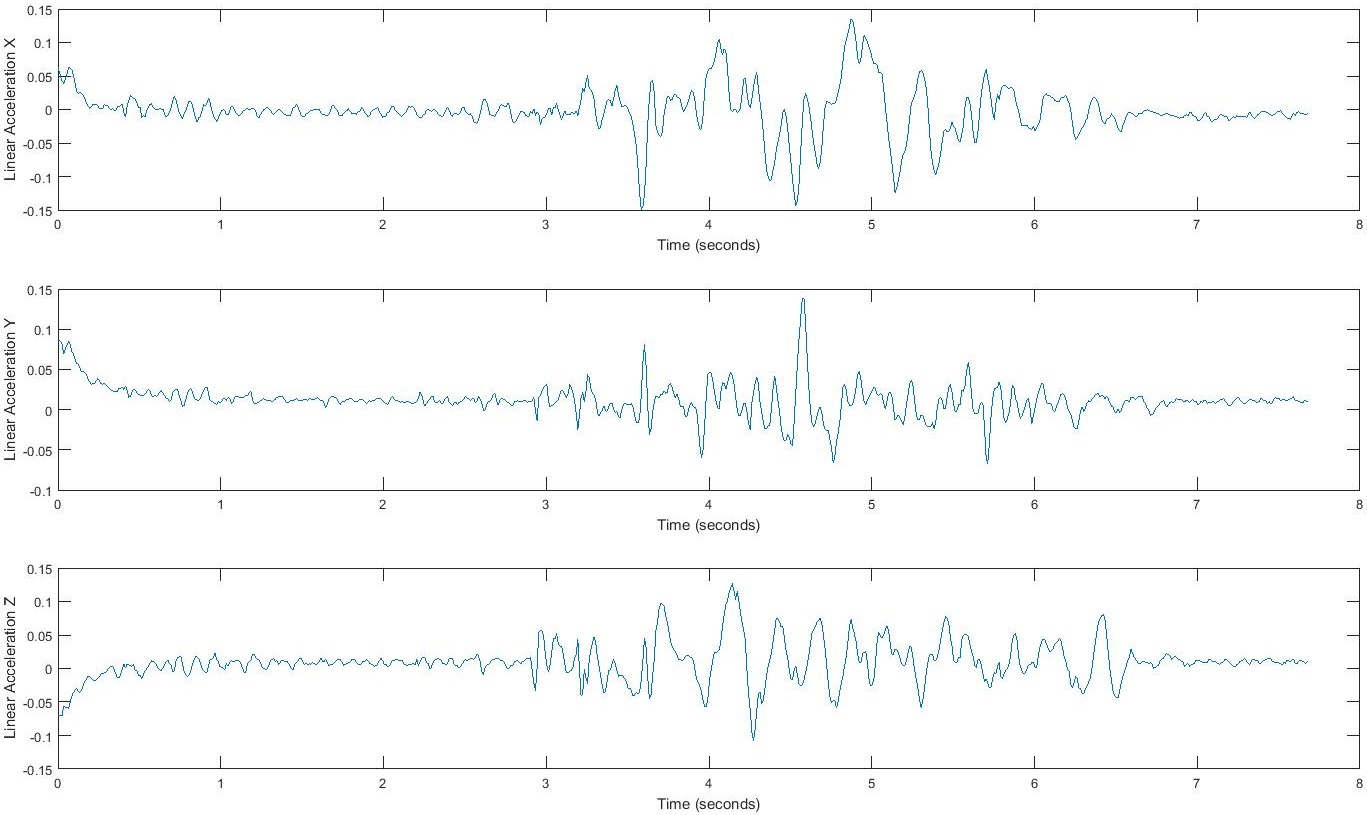}
\caption{Tandem gait linear acceleration signal}
\label{fig2}
\end{center}
\end{figure}

\begin{figure}[H]
\begin{center}
\includegraphics[scale=.295]{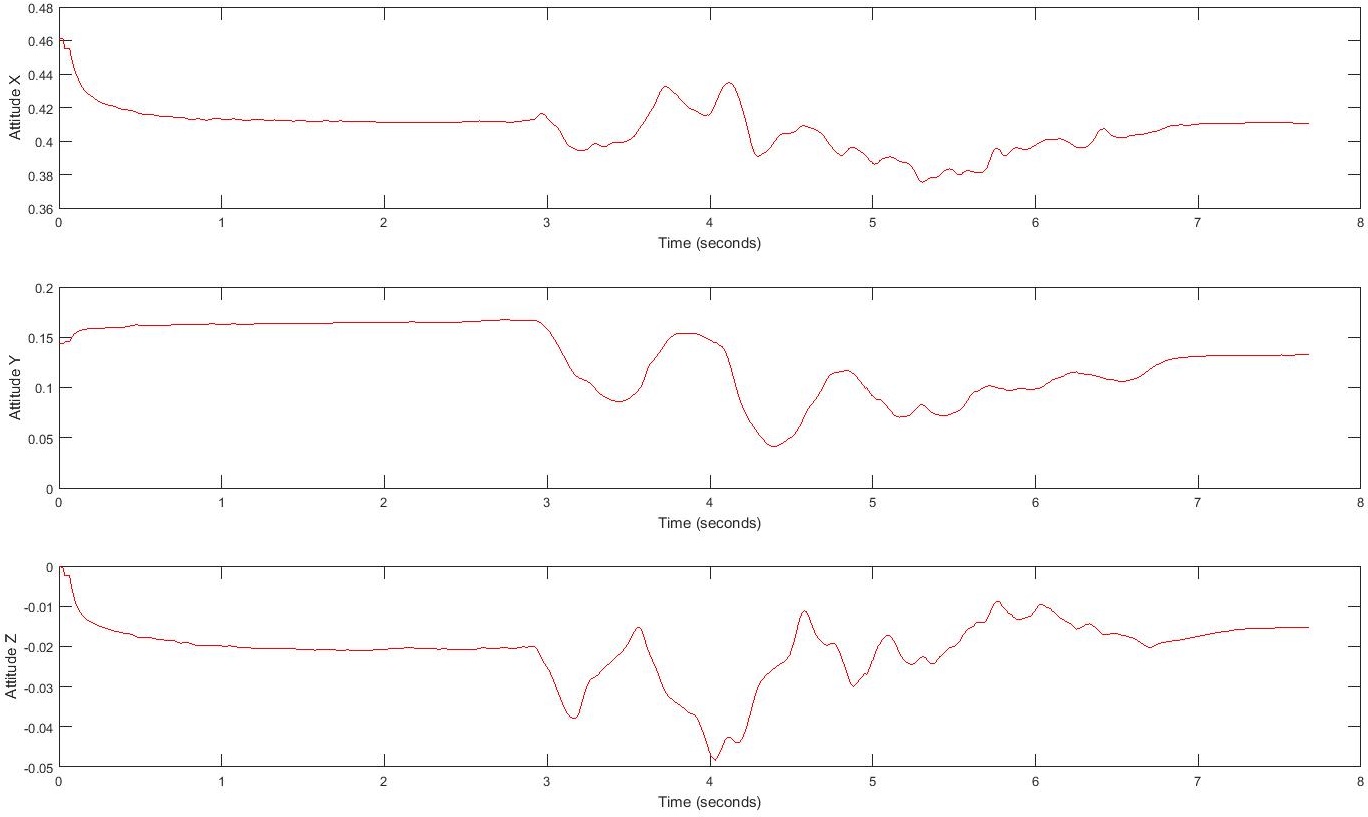}
\caption{Tandem gait linear attitude signal}
\label{fig3}
\end{center}
\end{figure}

In our experiment, subjects' movement generates linear acceleration and attitude signals measured by the smartphone. Raw acceleration data incorporates both gravity and the acceleration of the device. We removed the effects of gravity to measure acceleration of the device called linear acceleration which is determined by enhancing the gravity measurements with sensor fusion. Thus, the best result for computing linear acceleration needs not only an accelerometer, but also a gyroscope. We also measured the attitude of the device, which is the computed device orientation using the accelerometer, magnetometer, and gyroscope. These values yield the Euler angles of the device. Figure \ref{fig2} shows linear acceleration signal for one subject and Figure \ref{fig3} shows a three-axis signal of device attitude for the same subject.

\section{BAC Regression with Movement Pattern and Gait Features}

\begin{figure}[H]
\begin{center}
\includegraphics[scale=.3]{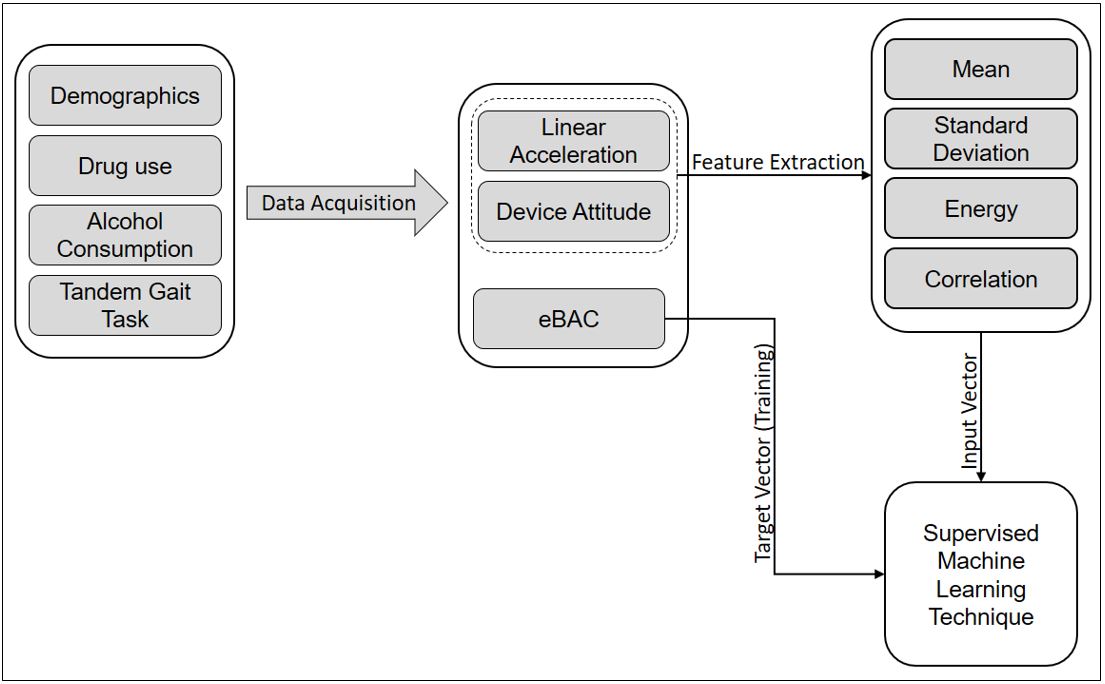}
\caption{Schematic diagram of the data flow for eBAC estimation}
\label{fig4}
\end{center}
\end{figure}

Using the sensor observations collected by the smartphone app (DrinkTRAC), inertial data are obtained during the tandem gait task. This set of data is used to explore gait and extract features to find the relationship between movement patterns and eBAC. We used these features as input vector into a supervised learning model that performs regression analysis to find the target value of eBAC. Figure \ref{fig4} shows a schematic diagram of the data flow for eBAC estimation. The procedures of data acquisition and feature extraction are explained in more detail in subsequent sections.

\subsection{Feature extraction for gait exploration}
\begin{figure}[H]
\begin{center}
\includegraphics[scale=.3]{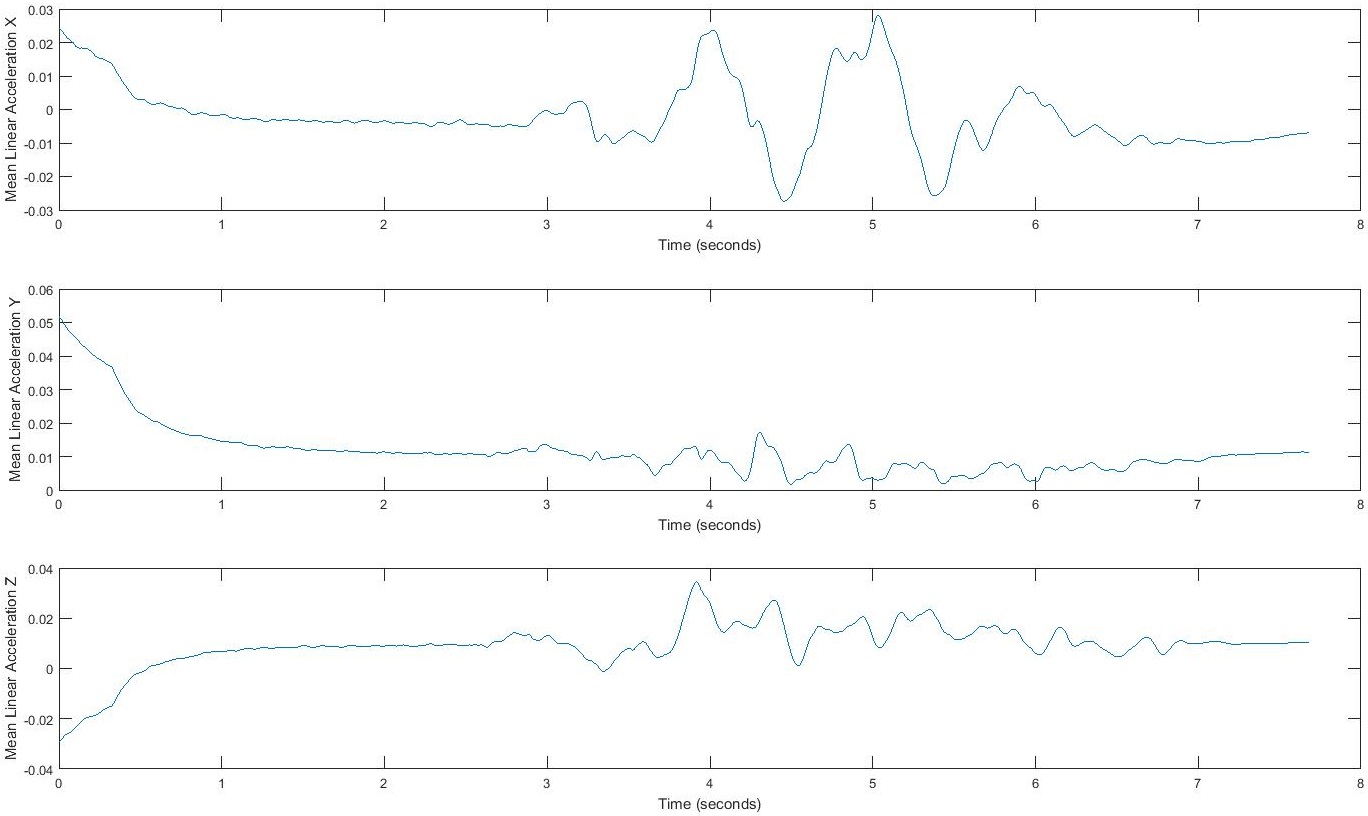}
\caption{Mean of linear acceleration, using a sliding window}
\label{fig5}
\end{center}
\end{figure}

\begin{figure}[H]
\begin{center}
\includegraphics[scale=.3]{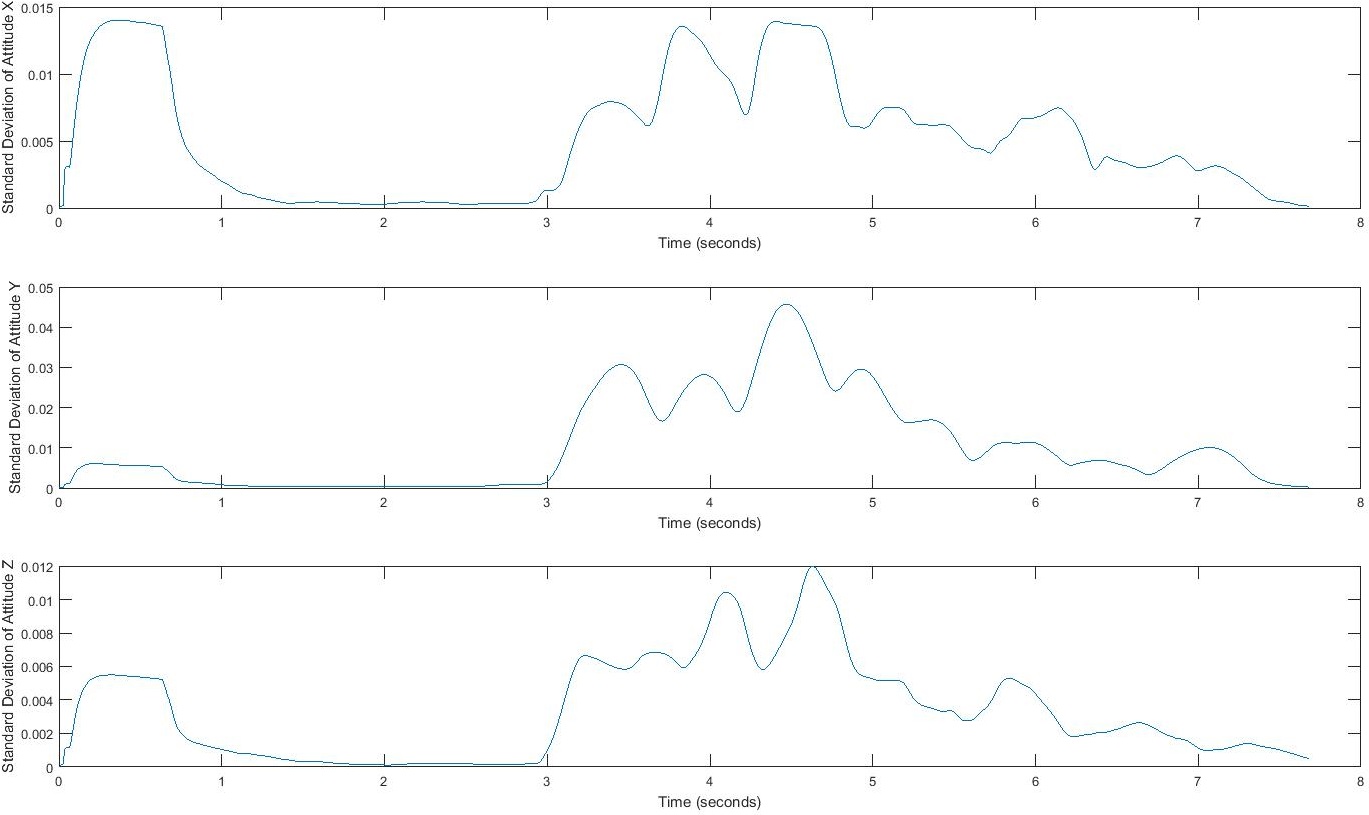}
\caption{Standard deviation of the attitude signal, computed using a sliding window}
\label{fig6}
\end{center}
\end{figure}

Acquired inertial data is noisy and less informative, rather than processed signals, which can reveal more information. Hence, we extract features that can describe the properties of each demonstrating gait signal. We consider four features, i.e., mean, standard deviation, correlation, and energy, by using a sliding window over signals. Extracted features belong to either the time domain or the frequency-domain. The first three features, mean, standard deviation, and correlation, are time domain features, and energy is from the frequency domain. All measurements are in three dimensions; thus, resulting in a total of 24 possible features. The efficiency of these features has been discussed in \citep{bao2004activity},\citep{ravi2005activity},\citep{Gharani2017}. Energy in the frequency domain is computed by using a fast Fourier transform (FFT), which converts the signal to frequency. Using the window size of 128 samples enables us to quickly calculate these parameters. In fact, energy feature is the sum of the squared discrete FFT coefficient magnitudes of the signal. The sum was divided by the window length of the window for normalization \citep{bao2004activity},\citep{ravi2005activity},\citep{Gharani2017}. If $x_1,x_2, \cdots$ are the FFT components of the window then $Energy= \frac{\sum\limits_{i=1}^{\left| w \right|}{\left|x_i \right|}^2}{\left| w \right|}$. Energy and mean values differ for each movement pattern. Also, correlation effectively demonstrates translation in one dimension. Figure \ref{fig5} and Figure \ref{fig6} show the extracted features for a gait signal.

\subsection{Bayesian regularized neural network (BRNN) for BAC regression}
Neural networks can model the relationship between an input vector x and an output y. The learning process involves adjusting the parameters in a way that enables the network to predict the output value for new input vectors. Neural networks are extensively used for slolving various types of clustering \citep{Gharani2015},\citep{Summers2014},\citep{gharani2014modeling}, classification \citep{Krizhevsky2012},and regression \citep{Ticknor2013} problems. One of the advantages of using neural networks for regression and predicting values is that it uses a nonlinear sigmoid function in a hidden layer, which enhances its computational flexibility, as compared to a standard linear regression model \citep{burden2009bayesian}. Thus, we applied a neural network model to estimate eBAC. In order to efficiently design and train a neural network, we must find an appropriate network architecture, determine a suitable training set, and compute the corresponding parameters of the network (such as weights and learning rate) by using an efficient and effective algorithm. In the rest of this section, we explain the overall system architecture and the training process of the parameter.

\subsection{Neural network architecture}

In this work, we used multilayer perceptron (MLP), a BRNN, to model the nonlinear relationships between input vectors, the extracted gait features, and the output (eBAC value), with nonlinear transfer functions. The basic MLP network is designed by arranging units in a layered structure, where each neuron in a layer takes its input from the output of the previous layer or from an external input. Figure \ref{fig8} shows a schematic diagram of our MLP structure. The transfer functions of the hidden layer in our feedforward network are a sigmoid function Eq.(2). Since we use MLP as a regression technique, we should produce reasonable output values that are outside the range of [-1,1]. Hence, in the output layer, we use a linear transfer function. Therefore, we may use this type of network as a general function approximator which approximates the eBAC as a function of gait features. The mathematical model to compute this is as follows, in Eq. (2-3):

\begin{equation}
F(\gamma)=\frac{1}{1+exp(-\gamma)}
\end{equation}

\begin{equation}
y=b_0+\sum_{j=1}^{H}w_j.F(\sum_{i=1}^{N}w_{ij}+b_j)
\end{equation}

where $x_i$ is the input, $w_{ij}$ is nonlinear weights that connect input neurons to hidden layer neurons, and $w_j$ linear weights that connect the hidden neurons with the output layer.

\subsection{Training}

Numerous training algorithms and learning rules have been proposed for setting the weights and parameters in neural networks; however, it is not possible to determine a global minimum solution. Therefore, training a network is one of the most crucial steps for neural network design. Backpropagation, which is basically a gradient descent optimization technique, is a standard and basic technique for training feedforward neural networks; however, it has some limitations, such as slow convergence, local search nature, overfitting data, and being overtrained, which can cause a loss of the network’s ability to correctly estimate the output \citep{burden2009bayesian}. As a result, the validation of the models can be problematic. Moreover, optimization of the network architecture is sometimes time-consuming. There are some modifications to the backpropagation, such as conjugate-gradient and Levenberg–Marquardt algorithms, that are faster than any variant of the backpropagation algorithm \citep{masters1995advanced},\citep{Mohanty2010}. The Levenberg–Marquardt algorithm is for minimizing a sum of squared error \citep{Roweis1996},\citep{Gavin2013} and to overcome some of the limitations in the standard backpropagation algorithm, such as an overfitting problem.

\begin{figure}[H]
\begin{center}
\includegraphics[scale=0.4]{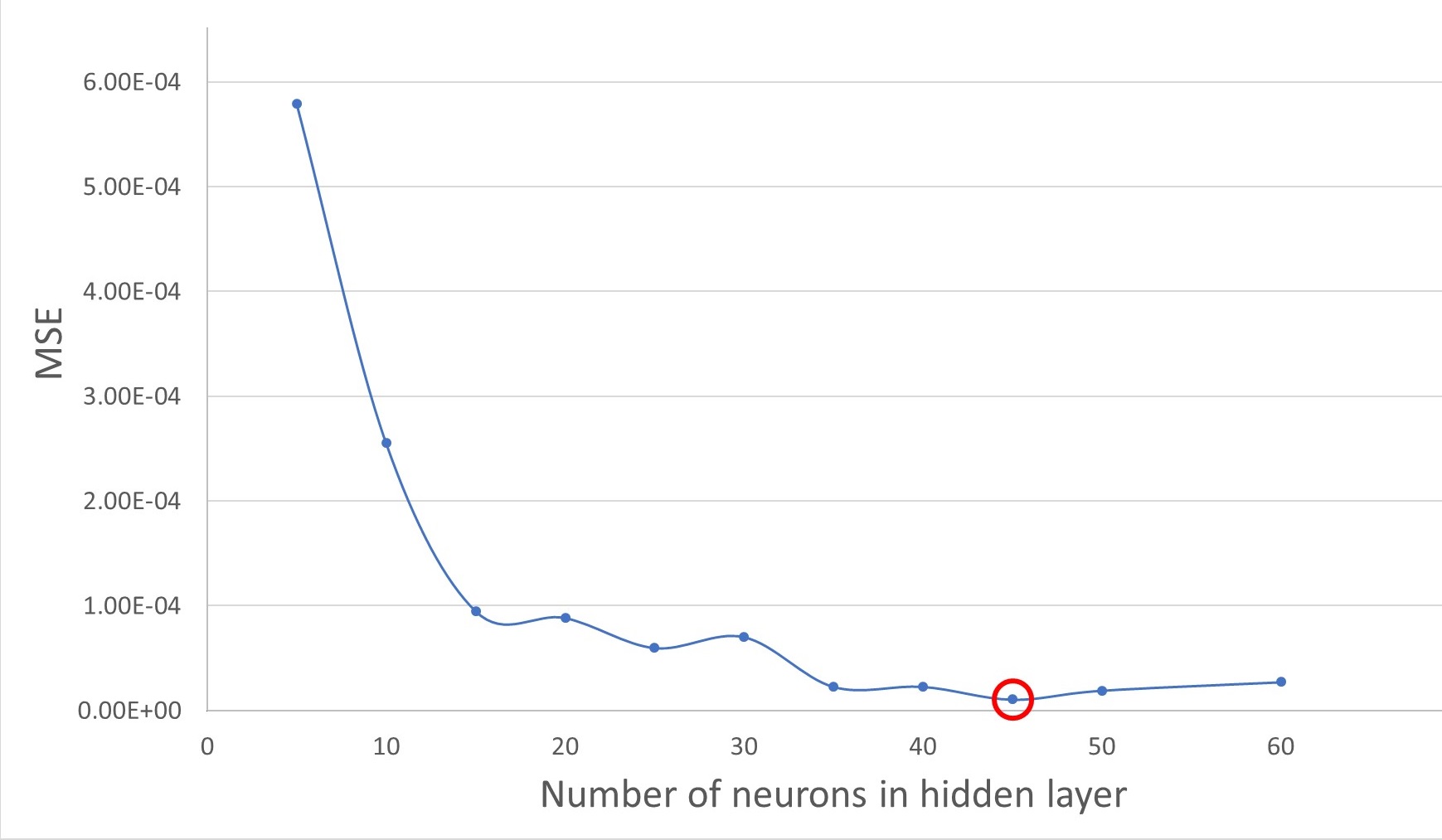}
\caption{Mean square error for differnt numbers of neurons in hidden layers}
\label{fig7}
\end{center}
\end{figure}

\begin{figure}[H]
\begin{center}
\includegraphics[scale=.8]{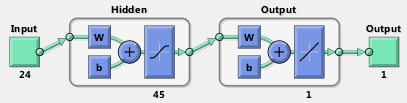}
\caption{The architecture of the feedforward neural network}
\label{fig8}
\end{center}
\end{figure}

Avoiding the overfitting problem in network architectures can be a serious challenge, because we try to achieve an accurate estimation of the modeled function by a neural network with a minimum number of input variables and parameters. Having too many neurons in the hidden layer can cause overfitting, since the noise of the data is modeled along with the trends. Furthermore, an insufficient number of neurons in the hidden layer can cause problems with the learning data. For the purpose of finding  the optimum number of neurons in the hidden layer, we conducted a model selection experiment with different number of neurons ranging from 5 to 60 and the cross-validation error for each setting was calculated. Figure \ref{fig7} presents the value of error for networks with one hidden layer where X-axis shows number of neurons and Y-axis represents error value. As it can be seen and highlighted, a hidden layer with 45 neurons is is a good fit for the dataset.

Using universal approximation theorem, it has been theoretically proven that a neural network with only one hidden layer using a bounded, continuous activation function can approximate any function \cite{hornik1993some},\citep{hornik1989multilayer}. Hence, in all configurations in the experiment and tests we only used one single hidden layer. Mackay \citep{MacKay1991} proposed a Bayesian regularization algorithm to meet such an overfitting challenge. Moreover, irrelevant and highly correlated parameters are another problem that can deteriorate the capability of the network to approximate the function, which can be solved by considering regularization \citep{burden2009bayesian}. Regularization can be modeled by incorporating Bayesian statistics. Through this method, we can remove most of the disadvantages of the feedforward neural network. In this study, we use a Bayesian regression neural network (BRNN) for the regression analysis. Thus, below we review this technique, which is a modification to the Levenberg-Marquardt algorithm.

\subsection{Levenberg-Marquardt algorithm}
The Levenberg-Marquardt algorithm is an iterative algorithm that finds the minimum of a multivariate function. It is the sum of squares of non-linear real-valued functions \citep{Lourakis2005}. Levenberg-Marquardt is widely used for solving non-linear least-squares problems and is usually considered as a standard technique for doing so. This algorithm is a curve-fitting method, a combination of gradient descent update and the Gauss-Newton update, two minimization methods. Eq. (4) represents gradient descent equations, and a normal equation for the Gauss-Newton update is shown by Eq. (5).

\begin{equation}
h_{gd}=\alpha \textbf{J}^{T}\textbf{W}(y-\hat{y})
\end{equation}

\begin{equation}
\textbf{J}^{T} \textbf{WJ}h_{gd}=\textbf{J}^{T}\textbf{W}(y-\hat{y})
\end{equation}

where:

\begin{equation}
\textbf{J}=
\left[
\begin{matrix}
\frac{\partial {e_1(w)}}{\partial w_1} & \frac{\partial {e_1(w)}}{\partial w_2} & \cdots & \frac{\partial {e_1(w)}}{\partial w_n}\\
\frac{\partial {e_2(w)}}{\partial w_1} & \frac{\partial {e_2(w)}}{\partial w_2} & \cdots & \frac{\partial {e_2(w)}}{\partial w_n}\\
\vdots &  \vdots & \ddots & \vdots \\
\frac{\partial {e_N(w)}}{\partial w_1} & \frac{\partial {e_N(w)}}{\partial w_2} & \cdots & \frac{\partial {e_N(w)}}{\partial w_n}\\
\end{matrix}
\right]
\end{equation}

As seen in Eq.(5), the Levenberg-Marquardt algorithm is a linear combination of the gradient descent update and the Gauss-Newton update, where the parameter updates adaptively vary between them. $\lambda$ determines this variation, and whenever the value of $\lambda$ is small, then it tends toward the Gauss-Newton update. Otherwise, when the $\lambda$ value is large, it will be closer to the gradient descent update. We started with a large $\lambda$ value, therefore the first updates were small values in the steepest-descent direction, just as the gradient descends.

\begin{equation}
\left[\textbf{J}^T\textbf{WJ}+\lambda \textbf{I}\right]h_{lm}=\textbf{J}^T\textbf{W}\left(y-\hat{y}\right)
\end{equation}

\subsection{Bayesian Regularization of Neural Networks}
Using a neural network for regression problems is preferable, as compared to other regression techniques. The first reason is the use of universal approximators, which can model any continuous nonlinear function \citep{MacKay1992}, although having appropriate training data is essential to this process. Nevertheless, it is likely to have both an overfitting and an overtraining problem. Overfitting behavior of the update function occurs in a way that causes it to decrease at the beginning as expected, but after overfitting the data, it starts to increase again. Therefore, the model overfits the data and generalizes poorly.

This problem is addressed by using the Bayesian approach, where the weights of the network are considered as random variables. It enables us to apply statistical techniques to estimate distribution parameters \citep{Foresee1997}. Furthermore, because Bayesian regularization considers not only the weight, but also the network structure as a probabilistic framework, it makes neural networks insensitive to the architecture of the network if a minimal architecture has been provided (Burden and Winkler 2008). In other words, Bayesian regularization can avoid overfitting by converting nonlinear systems into "well posed" problems \citep{burden2009bayesian},\citep{MacKay1992}. In conventional training, an optimal set of weights is chosen by minimizing the sum squared error of the model output and target value; in the Bayesian regularization, one more term is added to the objective function

\begin{equation}
F=\beta E_D\left(D|w,M\right)+\alpha E_w\left(w|M\right)
\end{equation}

where $E_D\left(D|w,M\right)=\frac{1}{N}\sum_{i,j}^{n}(\hat{y}_i-y_i)^2$  is the sum of squared errors, and $E_w=1/n\sum_{i,j}^{n}w_{ij}^2$ , which is called weight decay, is sum of square of the weights in the network. $\alpha$, the decay rate, and $\beta$ are the objective function parameters \citep{MacKay1992}. Considering the objective function in Eq. (8) and according to Bayes's rule, the posterior distribution of the neural network weights can be written as in \citep{Kayri2016}:
\begin{equation}
P\left(\textbf{w}|D,\alpha ,\beta ,M  \right)=\frac{P\left( D|\textbf{w},\beta ,M \right)P\left(\textbf{w}|\alpha,M \right)}{P\left(D|\alpha,\beta,M \right)}
\end{equation}

where $D$ is the dataset, $M$ is the network, and $\textbf{w}$ is the weight vector. Also, $P\left(\textbf{w}|\alpha,M \right)$ is the prior density, which represents our knowledge of the weights before any data is collected. $P\left( D|\textbf{w},\beta ,M \right)$ is the likelihood function, which is the probability of the data occurring, given weights w. The denominator of Eq.(9) is a normalization factor, which makes the summation of all probability 1 \citep{Foresee1997}.

\begin{equation}
P\left(D|\textbf{w},\beta,M\right)=\frac{1}{Z_D\left(\beta\right)}e^{-\beta E_D}
\end{equation}

\begin{equation}
P\left(\textbf{w}|\alpha,M\right)=\frac{1}{Z_w\left(\alpha\right)}e^{-\alpha E_w}
\end{equation}

where:
\begin{equation}
Z_w\left(\alpha\right)={\left(\frac{\pi}{\alpha}\right)}^{\frac{m}{2}}\qquad
Z_D\left(\beta\right)={\left(\frac{\pi}{\beta}\right)}^{\frac{n}{2}}
\end{equation}

so we have:
\begin{equation}
P\left(\textbf{w}|D,\alpha ,\beta ,M  \right)= \frac{\frac{1}{Z_D\left(\beta\right)}e^{-\beta E_D}\frac{1}{Z_w\left(\alpha\right)}e^{-\alpha E_w}}{C}=\frac{1}{Z_F\left(\alpha,\beta\right)}e^{-F\left(\textbf{w}\right)}
\end{equation}

Foresee and Hagan \cite{Foresee1997} demonstrated that maximizing the posterior probability $P\left(\textbf{w}|D,\alpha ,\beta ,M  \right)$ is equivalent to minimizing the regularized objective function $F=\beta E_D \left(D|w,M \right) +\alpha E_w\left(w|M \right) $. By using Bayes’s rule, the objective function parameters are optimized as follows \cite{Foresee1997}:

\begin{equation}
P\left(\alpha,\beta|D,M\right)=\frac{P\left(D|\alpha,\beta,M\right)P\left(\alpha,\beta|M\right)}{P\left(D|M\right)}
\end{equation}

\begin{equation}
P\left(D|\textbf{w},\alpha,M\right)=\frac{1}{Z_\textbf{w}\left(\alpha\right)}e^{-\alpha E_D}
\end{equation}
\begin{equation}
P\left(\textbf{w}|\beta,M\right)=\frac{1}{Z_\textbf{w}\left(\beta\right)}e^{-\beta E_\textbf{w}}
\end{equation}

\begin{equation}
P\left(\textbf{w}|D,\alpha ,\beta ,M  \right)=\frac{1}{Z_F\left(\alpha,\beta\right)}e^{-F\left(\textbf{w}\right)}
\end{equation}

\begin{equation}
P\left(D|\alpha,\beta,M\right)=\frac{P\left(D|\textbf{w},\beta,M\right)P\left(\textbf{w}|\alpha,M\right)}{P\left(\textbf{w}|D,\alpha,\beta,M\right)}=\frac{Z_F\left(\alpha,\beta\right)}{Z_D\left(\beta\right)Z_\textbf{w}\left(\alpha\right)}
\end{equation}

\begin{equation}
Z_F\left(\alpha,\beta\right)\approx {|\textbf{H}^{MAP}|}^{\frac{1}{2}}e^{-F\left(\textbf{w}^{MAP}\right)}
\end{equation}

where $\textbf{H}^{MAP}$ is the Hessian matrix ($\textbf{H}=\beta \nabla ^2E_D +\alpha \nabla ^2 E_w$) of the objective function, and $MAP$ stands for maximum a posteriori. By substituting the H matrix, we are able to solve for the optimal values for $\alpha$ and $\beta$:

\begin{equation}
\alpha^{MP}=\frac{\gamma}{2E_w\left(\textbf{w}^{MP}\right)}
\end{equation}

\begin{equation}
\beta^{MP}=\frac{n-\gamma}{2E_D\left(\textbf{w}^{MP}\right)}
\end{equation}

where $\gamma$ is the effective number of parameters and calculated as $\gamma=N-2\alpha^{MP}tr\left(\textbf{H}^{MP}\right)^{-1}$ and N is the total number of parameters in the network. Hessian matrix of F(w) must be computed; however, Foresee and Hagan \cite{Foresee1997} proposed using the Levenberg-Marquardt optimization algorithm to find the minimum point.

\section{Results and Validation}
In this section, we evaluate the designed network and present the results. First, we compare different training algorithms; then we conduct an experiment to see how effective an artificial neural network would be when applied to this dataset, and compared the BRNN to Support vector machine (SVM) and linear regression.

\subsection{Comparison of different training algorithms}
We compared three training algorithms to show that BRNN approach is the most effective. Table \ref{table2} shows the result of mean squared error (MSE) and R values for conjugated gradient, Levenberg-Marquardt, and Bayesian regularization. Although the results of MSE and R for the training data set are close, using an independent data set that was separate from all the computations and testing the networks reveals that Bayesian regularization outperforms two other algorithms. Moreover, we prefer to use Bayesian regularization, since it also adjusts the effective parameters and influences the architecture of the network. Table \ref{table2} and Table \ref{table3} show the results for comparing different training and optimization algorithms.

\begin{table}[]
\centering
\caption{Testing training algorithms with testing data}
\label{table2}
\begin{tabular}{ccc}
\hline
\textbf{Training algorithm} & \textbf{MSE} & \textbf{R} \\ \hline
Conjugate-gradient          & 5.80e-04    & 0.883389 \\
Levenberg-Marquardt         & 1.90e-05    & 0.996381 \\
Bayesian regularization     & 5.09e-06    & 0.999034 \\ \hline
\end{tabular}
\end{table}

\begin{table}[]
\centering
\caption{Testing training algorithms with independent samples}
\label{table3}
\begin{tabular}{ccc}
\hline
\textbf{Training algorithm} & \textbf{MSE} & \textbf{R} \\ \hline
Conjugate-gradient          & 5.89e-04   & 0.881084 \\
Levenberg-Marquardt         & 3.06e-05   & 0.994294 \\
Bayesian regularization     & 1.06e-05   & 0.998056 \\ \hline
\end{tabular}
\end{table}

\subsection{Performance}
\begin{figure}[H]
\begin{center}
\includegraphics[width=\linewidth]{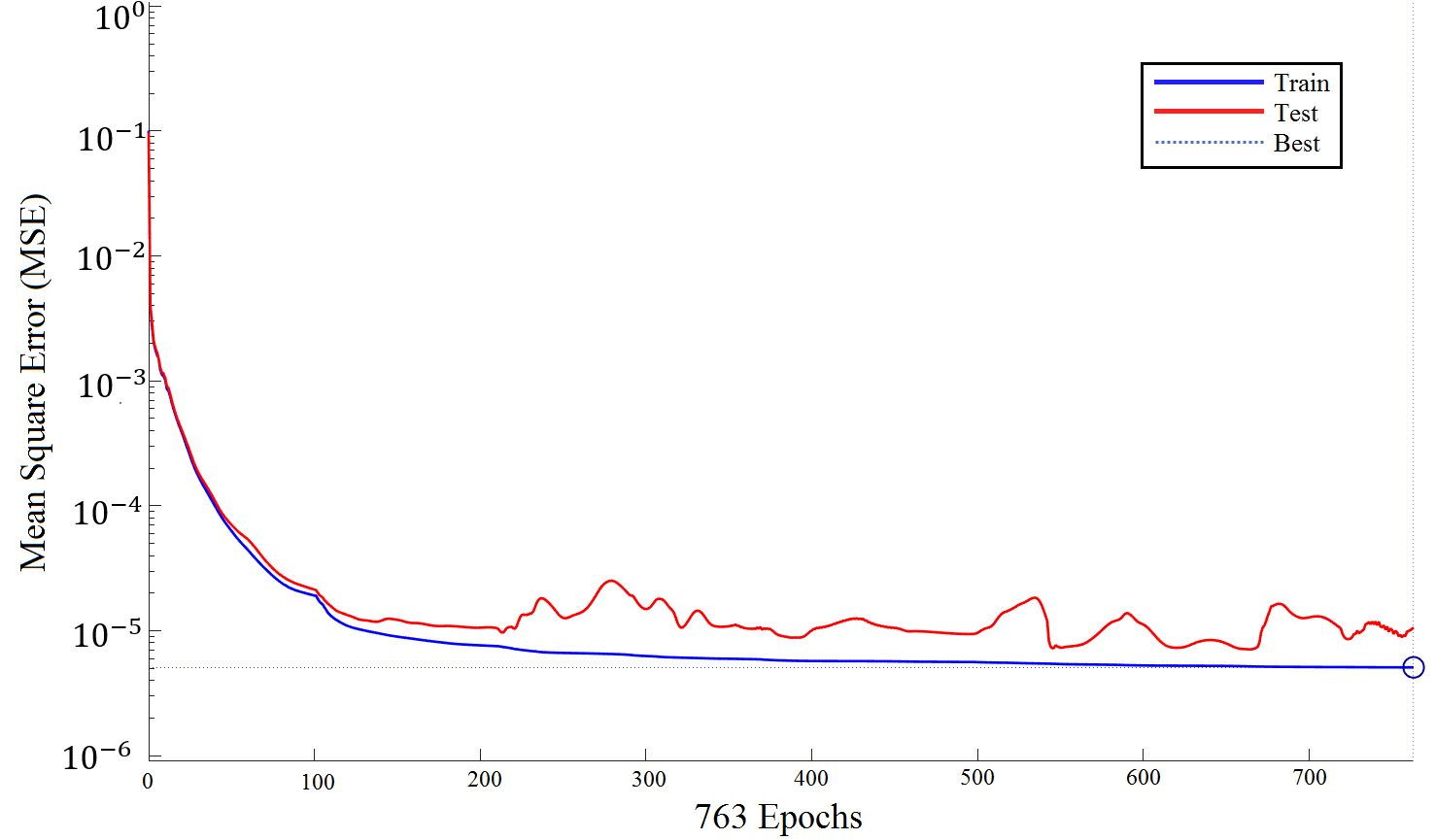}
\caption{Bayesian regularized neural network(BRNN) performance}
\label{fig9}
\end{center}
\end{figure}

Figure \ref{fig9} shows how network performance is improved during the training procedure. We measured the MSE for each of the training and test data sets. The BRNN algorithm does not use validation data.

\subsection{Error Histogram}
\begin{figure}[h]
\begin{center}
\includegraphics[scale=.25]{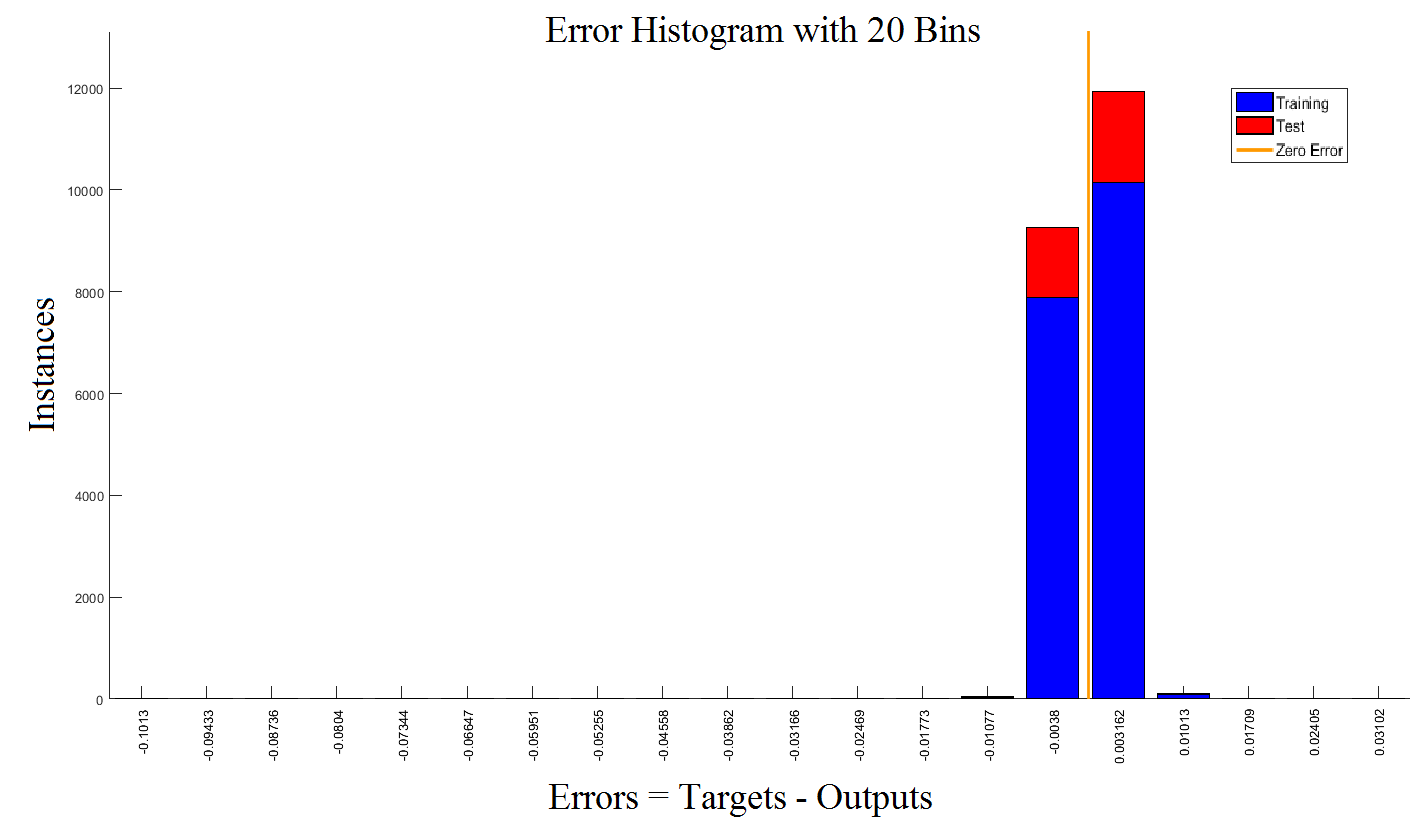}
\caption{Error histogram visualized errors between target values and predicted values after training a feedforward neural network with 20 bins}
\label{fig10}
\end{center}
\end{figure}

The blue bars represent training data and the red bars represent testing data (Figure \ref{fig10}). The histogram can give an indication of outliers, which are data points where the fit is significantly worse than that of most of the data. In this case, we can see that while most errors fall between -0.012 and +0.012, there are some training points and just a few test points that are outside of that range. These outliers are also visible on the testing regression plot (Figure \ref{fig11}). If the outliers are valid data points but are unlike the rest of the data, then the network is extrapolating for these points. It means more data similar to the outlier points should be considered in training analysis and that the network should be retrained.

\subsection{Regression results}
\begin{figure*}[t!]

	\begin{center}
		\begin{multicols}{3}
		
    		\includegraphics[width=\linewidth]{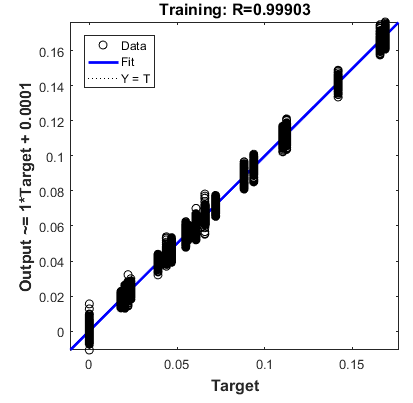}\par 
    		\subcaption{}
    		\includegraphics[width=\linewidth]{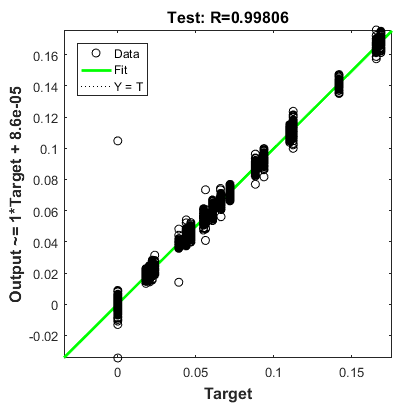}\par 
    		\subcaption{}
    		\includegraphics[width=\linewidth]{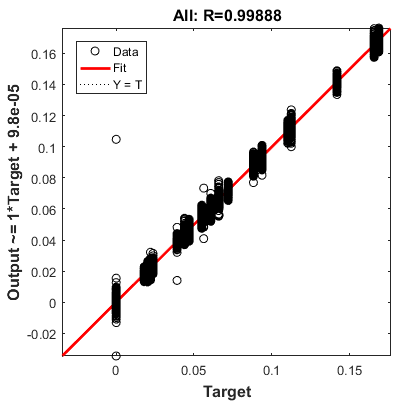}\par 
    		\subcaption{}
    	\end{multicols}
	\caption{The three plots represent the training, test, and all data}
	\label{fig11}
	\end{center}

\end{figure*}

The plots in Figure \ref{fig11} demonstrate the training, testing, and all data. There is a dashed line in each plot that represents the perfect result – outputs = targets, which can be seen on the regression diagrams. The solid line in each plot represents the best fit linear regression line between outputs and targets. On top of each plot, we also mentioned the R value as an indication of the relationship between the outputs and the targets. If R = 1, this indicates that there is an exact linear relationship between the outputs and the targets. If R is close to zero, then there is no linear relationship between the outputs and the targets. In Figure \ref{fig11}, as well as Table \ref{table2} and Table \ref{table3}, the R values show a reliable fit. The test results also show R values that are greater than 0.9. However, the scatter plot is helpful in showing that certain data points have poor fits.

\subsection{Comparison with other regression techniques}
We compared MLP with SVM and linear regression. We evaluated these techniques using the extracted features. We calculated their correlation coefficient, their mean absolute error, their root mean squared error, their relative absolute error, and their root relative squared error. Table \ref{table4} shows the results of the comparisons. Based on these results, and most importantly on the root mean squared error, MLP outperforms SVM and linear regression.
\begin{table}[H]
\centering

\caption{Comparison of different regression techniques for BAC estimation}
\label{table4}
\resizebox{\columnwidth}{!}{%
\begin{tabular}{cccccc}
\hline
Regression,technique & Correlation coefficient & Mean absolute error & Root mean squared error & Relative absolute error & Root relative squared error \\ \hline
MLP & 0.9009 & 0.0174 & 0.0226 & 40.6458 \% & 43.8853 \% \\
SVM & 0.3939 & 0.0362 & 0.0482 & 84.5348 \% & 93.6504 \% \\
Linear,Regression & 0.4367 & 0.0378 & 0.0463 & 88.2747 \% & 89.9583 \% \\ \hline
\end{tabular}%
}
\end{table}

\section{Related Work}
Smartphone-based alcohol consumption detection that evaluates a gait pattern captured by inertial sensors was proposed by \citep{Kao2012}, which labeled each gait signal with a Yes or a No in relation to alcohol intoxication. The study by Kao and colleagues (2012) did not examine the quantity of drinks consumed, but focused analyses solely on classifying a subject as intoxicated or not, thus limiting applicability across different ranges of BAC. Park et al. \citep{park2017unobtrusive} used a machine learning classifier to distinguish sober walking and alcohol-impaired walking by measuring gait features from a shoe-mounted accelerometer, which is impractical to use in the real world. Arnold et al. \citep{Arnold2015} also used smartphone inertial sensors to determine the number of drinks (not BAC), an approach which could be prone to errors given that the association between number of drinks and BACs varies by sex and weight.

Kao et. al. (2012) conducted a gait anomaly detection analysis by processing acceleration signals. Arnold et al. (2015) utilized naïve Bayes, decision trees, SVMs, and random Forest methods, where random Forest turned out to be the best classifier for their task. Also, in Virtual Breathalyzer \citep{nassi2016virtual} AdaBoost, gradient boosting, and decision trees were used for classifying whether the subject was intoxicated (yes or no). Furthermore, they implemented AdaBoost regression and regression trees (RT), as well as Lasso for estimation of BrAC.
Our approach differs from prior work in several ways. First, we use a widely available software platform for collecting movement data (Apple ResearchKit). Second, we calculate eBAC using established formulas, thus providing a more accurate representation of actual blood alcohol content than drink counting alone.  Third, we standardized the gait task, thus removing random variability in naturalistic walking. Fourth, we use not only accelerometers to understand movement data, but also consider gyroscope and magnetometer measurements.  Fifth, we use a sliding window technique for extracting features and feeding MLP, which outperformed the other evaluated approaches. Finally, instead of simply modeling association of gait with drinking (yes/no values), we examine these relationships across a range of BAC values.

\section{Conclusion and Future Direction}
This work provides initial support for the utility of using movement analysis in the real world to detect alcohol intoxication in terms of eBAC. We designed and used a smartphone application (DrinkTRAC) for collecting both self-reported alcohol consumption data to calculate eBAC and smartphone sensor-based movement data during a tandem gait task to understand gait impairments. We processed the raw data and extracted some features using a sliding window. These features were fed into a Bayesian regularized neural network, and we were able to model and fit a curve to the function of eBAC with movement pattern. The results indicated that the approach is reliable and that it can be used to identify the level of blood alcohol content during naturalistic drinking occasions. However, there are some limitations.

Some other aspects of body movement such as body sway can also be used for detecting intoxication where these aspects can be captured using the same phone sensor data. In order to detect not only new aspects of body movement, also improving the efficiency of the current model, it is needed to study and evaluate some other sophisticated gait features such as THD and Harmonic Distortion to select the best feature set for achieving better results. These various body movement study and gait-related feature selection can be considered as a possible direction for future research.

This pilot study is limited by its small sample size (n=10) and by the amount of missing EMA data (~70\%), particularly for the descending limb of alcohol intoxication. Nevertheless, as a proof-of-concept study,  it demonstrates the potential of accurate detection of drinking episodes using phone sensor data in the natural environment. Our findings may not be applicable to other populations, such as young adults with lighter alcohol use, or to other age groups, such as adolescents. The majority of participants were female and white, which limits our study’s overall generalizability. Therefore, there is a need to replicate the model in larger samples as future work.. Furtheremore, individual differences in tolerance to alcohol, which were not examined in this study, might affect the accuracy of the model, and warrant future research. Moreover, due to the importance of testing the relationship between BAC and gait in a more controlled environment, a crucial next step is to conduct a similar study while more tightly controlling alcohol intake.

Since participants should refrain from any non-drinking substance use during the sampling days, it is another source of limitation that in-the-moment data on other substance use was not collected, and there was no self-report or objective verification of other substance use, which might have affected eBAC and model accuracy. Additionally, the DrinkTRAC app was made only for iOS devices, which affected study eligibility, and limits the generalizability of results to other mobile devices. Self-reporting of alcohol use using EMA has demonstrated reliability and validity \citep{piasecki2012low}, but may be subject to bias. Future work could use transdermal alcohol sensors to validate findings and EMA schedule flexibility to reduce missing data.

\section*{Acknowledgment} The study was approved by the University of Pittsburgh IRB.





\section*{\refname}

\begin{thebibliography}{10}
\expandafter\ifx\csname url\endcsname\relax
  \def\url#1{\texttt{#1}}\fi
\expandafter\ifx\csname urlprefix\endcsname\relax\def\urlprefix{URL }\fi
\expandafter\ifx\csname href\endcsname\relax
  \def\href#1#2{#2} \def\path#1{#1}\fi

\bibitem{NCSAnalysis2015}
{National Center for Statistics and Analysis},
  \href{http://www-nrd.nhtsa.dot.gov/Pubs/812231.pdf}{{Alcohol- impaired
  driving: 2014 data}}, Tech. rep., National Highway Traffic Safety
  Administration, Washington, DC (December 2015).
\newline\urlprefix\url{http://www-nrd.nhtsa.dot.gov/Pubs/812231.pdf}

\bibitem{christoforou2013reaction}
Z.~Christoforou, M.~G. Karlaftis, G.~Yannis, {Reaction times of young
  alcohol-impaired drivers}, Accident Analysis {\&} Prevention 61 (2013)
  54--62.

\bibitem{steele1990alcohol}
C.~M. Steele, R.~A. Josephs, {Alcohol myopia: Its prized and dangerous
  effects.}, American Psychologist 45~(8) (1990) 921.

\bibitem{morris2014perceived}
D.~H. Morris, H.~R. Treloar, M.~E. Niculete, D.~M. McCarthy, {Perceived danger
  while intoxicated uniquely contributes to driving after drinking},
  Alcoholism: clinical and experimental research 38~(2) (2014) 521--528.

\bibitem{shults2001reviews}
R.~A. Shults, R.~W. Elder, D.~A. Sleet, J.~L. Nichols, M.~O. Alao, V.~G.
  Carande-Kulis, S.~Zaza, D.~M. Sosin, R.~S. Thompson, T.~F. {on Community
  Preventive Services}, Others, {Reviews of evidence regarding interventions to
  reduce alcohol-impaired driving}, American journal of preventive medicine
  21~(4) (2001) 66--88.

\bibitem{nieschalk1999effects}
M.~Nieschalk, C.~Ortmann, A.~West, F.~Schm{\"{a}}l, W.~Stoll, G.~Fechner,
  {Effects of alcohol on body-sway patterns in human subjects}, International
  journal of legal medicine 112~(4) (1999) 253--260.

\bibitem{jansen1985gait}
E.~C. Jansen, H.~H. Thyssen, J.~Brynskov, {Gait analysis after intake of
  increasing amounts of alcohol}, International Journal of Legal Medicine
  94~(2) (1985) 103--107.

\bibitem{anderson2015pew}
M.~Anderson, L.~Rainie, {Pew Research Center}, Technology Device Ownership:
  2015 29.

\bibitem{Arnold2015}
Z.~Arnold, D.~Larose, E.~Agu,
  \href{http://ieeexplore.ieee.org/document/7349720/}{{Smartphone Inference of
  Alcohol Consumption Levels from Gait}}, in: 2015 International Conference on
  Healthcare Informatics, IEEE, 2015, pp. 417--426.
\newblock \href {http://dx.doi.org/10.1109/ICHI.2015.59}
  {\path{doi:10.1109/ICHI.2015.59}}.
\newline\urlprefix\url{http://ieeexplore.ieee.org/document/7349720/}

\bibitem{Aiello2016}
Aiello, Agu, \href{http://ieeexplore.ieee.org/document/7764559/}{{Investigating
  postural sway features, normalization and personalization in detecting blood
  alcohol levels of smartphone users}}, in: 2016 IEEE Wireless Health (WH),
  IEEE, 2016, pp. 1--8.
\newblock \href {http://dx.doi.org/10.1109/WH.2016.7764559}
  {\path{doi:10.1109/WH.2016.7764559}}.
\newline\urlprefix\url{http://ieeexplore.ieee.org/document/7764559/}

\bibitem{bae2017mobile}
S.~Bae, T.~Chung, D.~Ferreira, A.~K. Dey, B.~Suffoletto, Mobile phone sensors
  and supervised machine learning to identify alcohol use events in young
  adults: Implications for just-in-time adaptive interventions, Addictive
  Behaviors.

\bibitem{winek1984blood}
C.~L. Winek, F.~M. Esposito,
  \href{http://www.ncbi.nlm.nih.gov/pubmed/3835425}{{Blood alcohol
  concentrations: factors affecting predictions.}}, Legal medicine (1985)
  34--61.
\newline\urlprefix\url{http://www.ncbi.nlm.nih.gov/pubmed/3835425}

\bibitem{greenfield2014biomonitoring}
T.~K. Greenfield, J.~Bond, W.~C. Kerr, {Biomonitoring for improving alcohol
  consumption surveys: the new gold standard?}, Alcohol research: current
  reviews 36~(1) (2014) 39.

\bibitem{wechsler2002trends}
H.~Wechsler, J.~E. Lee, M.~Kuo, M.~Seibring, T.~F. Nelson, H.~Lee, {Trends in
  college binge drinking during a period of increased prevention efforts:
  Findings from 4 Harvard School of Public Health College Alcohol Study
  surveys: 1993--2001}, Journal of American college health 50~(5) (2002)
  203--217.

\bibitem{widmark1932theoretischen}
E.~M.~P. Widmark, {Die theoretischen Grundlagen und die praktische
  Verwendbarkeit der gerichtlich-medizinischen Alkoholbestimmung}, Urban {\&}
  Schwarzenberg, 1932.

\bibitem{hustad2005using}
J.~T.~P. Hustad, K.~B. Carey, {Using calculations to estimate blood alcohol
  concentrations for naturally occurring drinking episodes: a validity study.},
  Journal of Studies on Alcohol 66~(1) (2005) 130--138.

\bibitem{babor2000talk}
T.~F. Babor, K.~Steinberg, R.~A.~Y. Anton, F.~{Del Boca}, {Talk is cheap:
  measuring drinking outcomes in clinical trials.}, Journal of studies on
  alcohol 61~(1) (2000) 55--63.

\bibitem{alessi2017experiences}
S.~M. Alessi, N.~P. Barnett, N.~M. Petry, {Experiences with SCRAMx alcohol
  monitoring technology in 100 alcohol treatment outpatients}, Drug and Alcohol
  Dependence 178 (2017) 417--424.

\bibitem{simons2015quantifying}
J.~S. Simons, T.~A. Wills, N.~N. Emery, R.~M. Marks, {Quantifying alcohol
  consumption: self-report, transdermal assessment, and prediction of
  dependence symptoms}, Addictive behaviors 50 (2015) 205--212.

\bibitem{alessi2013randomized}
S.~M. Alessi, N.~M. Petry, {A randomized study of cellphone technology to
  reinforce alcohol abstinence in the natural environment}, Addiction 108~(5)
  (2013) 900--909.

\bibitem{suffoletto2017using}
B.~Suffoletto, P.~Gharani, T.~Chung, H.~Karimi, {Using Phone Sensors and an
  Artificial Neural Network to Detect Gait Changes During Drinking Episodes in
  the Natural Environment}, Gait {\&} Posture.

\bibitem{leffingwell2013continuous}
T.~R. Leffingwell, N.~J. Cooney, J.~G. Murphy, S.~Luczak, G.~Rosen, D.~M.
  Dougherty, N.~P. Barnett, {Continuous objective monitoring of alcohol use:
  twenty-first century measurement using transdermal sensors}, Alcoholism:
  Clinical and Experimental Research 37~(1) (2013) 16--22.

\bibitem{karns2016time}
T.~E. Karns-Wright, J.~D. Roache, N.~Hill-Kapturczak, Y.~Liang, J.~Mullen,
  D.~M. Dougherty, {Time Delays in Transdermal Alcohol Concentrations Relative
  to Breath Alcohol Concentrations.}, Alcohol and alcoholism (Oxford,
  Oxfordshire) 52~(1) (2017) 35--41.
\newblock \href {http://dx.doi.org/10.1093/alcalc/agw058}
  {\path{doi:10.1093/alcalc/agw058}}.

\bibitem{muraven2005daily}
M.~Muraven, R.~L. Collins, S.~Shiffman, J.~A. Paty, {Daily fluctuations in
  self-control demands and alcohol intake.}, Psychology of Addictive Behaviors
  19~(2) (2005) 140.

\bibitem{us2005helping}
N.~I. on~Alcohol~Abuse, A.~N. Publication, {Helping patients who drink too
  much: a clinician's guide}, Tech. Rep. 07-3769, National Institutes of Health
  (2005).

\bibitem{shiffman2009ecological}
S.~Shiffman, {Ecological momentary assessment (EMA) in studies of substance
  use.}, Psychological assessment 21~(4) (2009) 486.

\bibitem{Wray2014}
T.~B. Wray, J.~E. Merrill, P.~M. Monti, {Using Ecological Momentary Assessment
  (EMA) to Assess Situation-Level Predictors of Alcohol Use and Alcohol-Related
  Consequences.}, Alcohol research : current reviews 36~(1) (2014) 19--27.

\bibitem{lucas2014s}
G.~M. Lucas, J.~Gratch, A.~King, L.-P. Morency, {It's only a computer: virtual
  humans increase willingness to disclose}, Computers in Human Behavior 37
  (2014) 94--100.

\bibitem{babor1992guidelines}
T.~F. Babor, J.~R. De~La~Fuente, J.~Saunders, M.~Grant, et~al., Audit the
  alcohol use disorders identification test: Guidelines for use in primary
  health care, PROGRAMME ON SUBSTANCE ABUSE.

\bibitem{del2004up}
F.~K. {Del Boca}, J.~Darkes, P.~E. Greenbaum, M.~S. Goldman, {Up close and
  personal: Temporal variability in the drinking of individual college students
  during their first year}, Journal of consulting and clinical psychology
  72~(2) (2004) 155--164.

\bibitem{matthews1979estimating}
D.~B. Matthews, W.~R. Miller, {Estimating blood alcohol concentration: Two
  computer programs and their applications in therapy and research}, Addictive
  behaviors 4~(1) (1979) 55--60.

\bibitem{suffoletto2017can}
B.~Suffoletto, A.~Goyal, J.~C. Puyana, T.~Chung, {Can an App Help Identify
  Psychomotor Function Impairments During Drinking Occasions in the Real World?
  A Mixed Method Pilot Study}, Substance Abuse.

\bibitem{bao2004activity}
L.~Bao, S.~Intille, Activity recognition from user-annotated acceleration data,
  Pervasive computing (2004) 1--17.

\bibitem{ravi2005activity}
N.~Ravi, N.~Dandekar, P.~Mysore, M.~L. Littman, {Activity recognition from
  accelerometer data}, in: AAAI, Vol.~5, 2005, pp. 1541--1546.

\bibitem{Gharani2017}
P.~Gharani, H.~Karimi, {Context-aware obstacle detection for navigation by
  visually impaired}, Image and Vision Computing 64.
\newblock \href {http://dx.doi.org/10.1016/j.imavis.2017.06.002}
  {\path{doi:10.1016/j.imavis.2017.06.002}}.

\bibitem{Gharani2015}
P.~Gharani, K.~Stewart, G.~Ryan, {An enhanced approach for modeling spatial
  accessibility for in vitro fertilization services in the rural Midwestern
  United States}, Applied Geography 64.
\newblock \href {http://dx.doi.org/10.1016/j.apgeog.2015.08.005}
  {\path{doi:10.1016/j.apgeog.2015.08.005}}.

\bibitem{Summers2014}
K.~Summers, K.~Stewart, P.~Gharani, G.~Ryan, B.~{Van Voorhis},
  \href{http://linkinghub.elsevier.com/retrieve/pii/S0015028214015660}{{Geospatial
  modeling of in-vitro fertilization (IVF) accessibility in a rural midwestern
  state}}, Fertility and Sterility 102~(3) (2014) e276--e277.
\newblock \href {http://dx.doi.org/10.1016/j.fertnstert.2014.07.939}
  {\path{doi:10.1016/j.fertnstert.2014.07.939}}.
\newline\urlprefix\url{http://linkinghub.elsevier.com/retrieve/pii/S0015028214015660}

\bibitem{gharani2014modeling}
P.~Gharani, {Modeling spatial accessibility for in-vitro fertility (IVF) care
  services in Iowa}, Master's thesis, University of Iowa (2014).

\bibitem{Krizhevsky2012}
A.~Krizhevsky, I.~Sutskever, G.~E. Hinton, Imagenet classification with deep
  convolutional neural networks, in: Advances in neural information processing
  systems, 2012, pp. 1097--1105.

\bibitem{Ticknor2013}
J.~L. Ticknor, \href{http://dx.doi.org/10.1016/j.eswa.2013.04.013}{{A Bayesian
  regularized artificial neural network for stock market forecasting}}, Expert
  Systems with Applications 40~(14) (2013) 5501--5506.
\newblock \href {http://dx.doi.org/10.1016/j.eswa.2013.04.013}
  {\path{doi:10.1016/j.eswa.2013.04.013}}.
\newline\urlprefix\url{http://dx.doi.org/10.1016/j.eswa.2013.04.013}

\bibitem{burden2009bayesian}
F.~Burden, D.~Winkler, Bayesian regularization of neural networks, Artificial
  Neural Networks: Methods and Applications (2009) 23--42.

\bibitem{masters1995advanced}
T.~Masters, {Advanced algorithms for neural networks: a C++ sourcebook}, John
  Wiley {\&} Sons, Inc., 1995.

\bibitem{Mohanty2010}
S.~Mohanty, M.~K. Jha, A.~Kumar, K.~P. Sudheer, {Artificial neural network
  modeling for groundwater level forecasting in a river island of eastern
  India}, Water Resources Management 24~(9) (2010) 1845--1865.
\newblock \href {http://dx.doi.org/10.1007/s11269-009-9527-x}
  {\path{doi:10.1007/s11269-009-9527-x}}.

\bibitem{Roweis1996}
S.~Roweis, {Levenberg-Marquardt Optimization}, Notes, University Of Toronto.

\bibitem{Gavin2013}
H.~P. Gavin, {The Levenberg-Marquardt method for nonlinear least squares
  curve-fitting problems}, Department of Civil and Environmental Engineering,
  Duke University (2013) 1--17\href
  {http://dx.doi.org/10.1080/10426914.2014.941480}
  {\path{doi:10.1080/10426914.2014.941480}}.

\bibitem{hornik1993some}
K.~Hornik, {Some new results on neural network approximation}, Neural networks
  6~(8) (1993) 1069--1072.

\bibitem{hornik1989multilayer}
K.~Hornik, M.~Stinchcombe, H.~White, {Multilayer feedforward networks are
  universal approximators}, Neural networks 2~(5) (1989) 359--366.

\bibitem{MacKay1991}
D.~J.~C. Mackay, {Bayesian Methods for Adaptive Models}, thesis, CIT (1991) 98.

\bibitem{Lourakis2005}
M.~I.~a. Lourakis, \href{http://www.ics.forth.gr/{~}lourakis/}{{A Brief
  Description of the Levenberg-Marquardt Algorithm Implemened by levmar}},
  Matrix 3 (2005) 2.
\newblock \href {http://dx.doi.org/10.1016/j.ijinfomgt.2009.10.001}
  {\path{doi:10.1016/j.ijinfomgt.2009.10.001}}.
\newline\urlprefix\url{http://www.ics.forth.gr/{~}lourakis/}

\bibitem{MacKay1992}
D.~J.~C. MacKay, {A Practical Bayesian Framework for Backpropagation Networks},
  Neural Computation 4~(3) (1992) 448--472.
\newblock \href {http://dx.doi.org/10.1162/neco.1992.4.3.448}
  {\path{doi:10.1162/neco.1992.4.3.448}}.

\bibitem{Foresee1997}
F.~D. Foresee, M.~T. Hagan, {Gauss-Newton approximation to Bayesian
  regularization}, Proceedings of the 1997 International Joint Conference on
  Neural Networks (1997) 1930--1935\href
  {http://dx.doi.org/10.1109/ICNN.1997.614194}
  {\path{doi:10.1109/ICNN.1997.614194}}.

\bibitem{Kayri2016}
M.~Kayri, \href{http://www.mdpi.com/2297-8747/21/2/20}{{Predictive Abilities of
  Bayesian Regularization and Levenberg–Marquardt Algorithms in Artificial
  Neural Networks: A Comparative Empirical Study on Social Data}}, Mathematical
  and Computational Applications 21~(2) (2016) 20.
\newblock \href {http://dx.doi.org/10.3390/mca21020020}
  {\path{doi:10.3390/mca21020020}}.
\newline\urlprefix\url{http://www.mdpi.com/2297-8747/21/2/20}

\bibitem{Kao2012}
H.-L.~C. Kao, B.-J. Ho, A.~C. Lin, H.-H. Chu,
  \href{http://dl.acm.org/citation.cfm?id=2370216.2370354}{{Phone-based gait
  analysis to detect alcohol usage}}, Proceedings of the 2012 ACM Conference on
  Ubiquitous Computing - UbiComp '12 (2012) 661\href
  {http://dx.doi.org/10.1145/2370216.2370354}
  {\path{doi:10.1145/2370216.2370354}}.
\newline\urlprefix\url{http://dl.acm.org/citation.cfm?id=2370216.2370354}

\bibitem{park2017unobtrusive}
E.~Park, S.~I. Lee, H.~S. Nam, J.~H. Garst, A.~Huang, A.~Campion, M.~Arnell,
  N.~Ghalehsariand, S.~Park, H.-j. Chang, et~al., Unobtrusive and continuous
  monitoring of alcohol-impaired gait using smart shoes, Methods of information
  in medicine 56~(1) (2017) 74--82.

\bibitem{nassi2016virtual}
B.~Nassi, L.~Rokach, Y.~Elovici, Virtual breathalyzer, arXiv preprint
  arXiv:1612.05083.

\bibitem{piasecki2012low}
T.~M. Piasecki, K.~J. Alley, W.~S. Slutske, P.~K. Wood, K.~J. Sher,
  S.~Shiffman, A.~C. Heath, {Low sensitivity to alcohol: relations with
  hangover occurrence and susceptibility in an ecological momentary assessment
  investigation}, Journal of studies on alcohol and drugs 73~(6) (2012)
  925--932.

\end{thebibliography}







\end{document}